\documentclass[fleqn,usenatbib]{mnras}

\usepackage{newtxtext,newtxmath}

\usepackage[T1]{fontenc}

\DeclareRobustCommand{\VAN}[3]{#2}
\let\VANthebibliography\thebibliography
\def\thebibliography{\DeclareRobustCommand{\VAN}[3]{##3}\VANthebibliography}

\usepackage{graphicx}	
\usepackage{amsmath}	
\usepackage{xspace}
\usepackage{siunitx}
\usepackage{subfig}
\usepackage{caption}

\DeclareCaptionFormat{cont}{#1 (cont.)#2#3\par}

\newcommand{\hi}{{\sc H\,i}\xspace} 

\newcommand{\msun}{{M$_\odot\,$}}

\newcommand{\kms}{$\,$km$\,$s$^{-1}$}

\newcommand{\perbeam}{beam$^{-1}$}

\title[HI group in MIGHTEE]{MIGHTEE-HI: Discovery of an \hi-rich galaxy group at $z=0.044$ with MeerKAT}

\author[S. Ranchod et al.]{Shilpa Ranchod$^{1,2}$\thanks{E-mail: shilparanchod@gmail.com},
Roger P. Deane,$^{1,2}$, Anastasia A. Ponomareva$^3$, Tariq Blecher$^4$, Bradley S. Frank$^{5,6,7}$, \newauthor Matt J. Jarvis$^{3,8}$, Natasha Maddox$^9$, Wanga Mulaudzi$^{5}$, Marcin Glowacki$^{8,10}$, Kelley M. Hess$^{11,12}$, \newauthor Madalina Tudorache$^3$, Lourdes Verdes-Montenegro$^{13}$,
Nathan J. Adams$^3$, Rebecca A. A. Bowler$^3$, 
\newauthor Jordan. D. Collier$^{7,14}$, Russ Taylor$^{7,10}$
\\
$^{1}$Department of Physics, University of Pretoria, Private Bag X20, Pretoria 0028, South Africa\\
$^{2}$Wits Centre for Astrophysics, School of Physics, University of the Witwatersrand, 1 Jan Smuts Avenue, 2000, South Africa\\
$^{3}$Astrophysics, University of Oxford, Denys Wilkinson Building, Keble Road, Oxford OX1 3RH, UK\\
$^{4}$Department of Physics and Electronics, Rhodes University, PO Box 94, Grahamstown 6140, Eastern Cape, South Africa\\
$^{5}$Department of Astronomy, University of Cape Town, Private Bag X3, Rondebosch 7701, South Africa \\
$^6$ South African Radio Astronomy Observatory, 2 Fir Street, Observatory, 7925, South Africa \\
$^7$ The Inter-University Institute for Data Intensive Astronomy (IDIA), Department of Astronomy, University of Cape Town, Private Bag X3, Rondebosch, \\ 
    7701, South Africa \\
$^{8}$ Department of Physics and Astronomy, University of the Western Cape, Robert Sobukwe Road, Bellville 7535, South Africa \\
$^9$ Faculty of Physics, Ludwig-Maximilians-Universit\"at, Scheinerstr. 1, 81679 Munich, Germany \\
$^{10}$ The Inter-University Institute for Data Intensive Astronomy (IDIA), Department of Physics and Astronomy, University of the Western Cape, Bellville \\
7535, South Africa \\
$^{11}$ ASTRON, the Netherlands Institute for Radio Astronomy, Postbus 2, 7990 AA, Dwingeloo, The Netherlands \\
$^{12}$ Kapteyn Astronomical Institute, University of Groningen, P.O. Box 800, 9700 AV Groningen, The Netherlands \\
$^{13}$ Instituto de Astrof\'isica de Andaluc\'ia (IAA-CSIC), Glorieta de la Astronom\'ia, 18008 Granada, Spain \\
$^{14}$ School of Science, Western Sydney University, Locked Bag 1797, Penrith, NSW 2751, Australia \\
}

\date{Accepted XXX. Received YYY; in original form ZZZ}

\pubyear{2021}

\begin{document}
\label{firstpage}
\pagerange{\pageref{firstpage}--\pageref{lastpage}}
\maketitle

\begin{abstract}
We present the serendipitous discovery of a galaxy group in the XMM-LSS field with MIGHTEE Early Science observations. Twenty galaxies are detected in \hi in this $z\sim0.044$ group, with a $3\sigma$ column density sensitivity of $N_\mathrm{HI} = 1.6\times10^{20}\,\mathrm{cm}^{-2}$. This group has not been previously identified, despite residing in a well-studied extragalactic legacy field. We present spatially-resolved \hi total intensity and velocity maps for each of the objects, which reveal environmental influence through disturbed morphologies. The group has a dynamical mass of $\log_{10}(M_\mathrm{dyn}/\mathrm{M}_\odot) = 12.32$, and is unusually gas-rich, with an \hi-to-stellar mass ratio of $\log_{10}(f_\mathrm{HI}^\mathrm{*}) = -0.2$, which is 0.7 dex greater than expected. The group's high \hi content, spatial, velocity, and identified galaxy type distributions strongly suggest that it is in the early stages of its assembly. The discovery of this galaxy group is an example of the importance of mapping spatially-resolved \hi in a wide range of environments, including galaxy groups. This scientific goal has been dramatically enhanced by the high sensitivity, large field-of-view, and wide instantaneous bandwidth of the MeerKAT telescope. 
\end{abstract}

\begin{keywords}
galaxies: evolution -- galaxies: groups: individual -- instrumentation: interferometer -- radio lines: galaxies
\end{keywords}



\section{Introduction}\label{sec:intro}
The current understanding of structure formation in the Universe is the $\Lambda$CDM hierarchical merging paradigm, where larger structures are formed by the mergers of smaller dark matter haloes \citep[e.g.][]{Springel_2017}. Galaxy groups, with typical virial masses of order $M_\mathrm{vir}\sim10^{12-14}\,$\msun \citep{Feldmann_2011}, occupy an interesting intermediary region within the dark matter halo mass spectrum to study the physical processes of galaxy evolution. In the local Universe, up to 50\% of galaxies reside in groups, most of which have only a handful of members \citep[$<10$;][]{Huchra_1982,Berlind_2006,Crook_2007,Lim_2017}. These tend to be spiral-galaxy dominated, and unlike clusters, do not host easily detectable X-ray emitting cores of hot intergalactic material, at least with current X-ray telescope sensitivity limits. Massive groups ($\sim 20$ members) are comparatively rare \citep[e.g.][]{Hess_2013}, but make up an important part of the continuum between low-mass groups and massive clusters. 

The halo mass, galaxy number density, and intergalactic medium density make groups an important laboratory of environmental transformation effects, and their context within a hierarchical galaxy formation and cosmological framework. Given their lower dark matter halo masses, the velocity dispersion for groups ranges between $\sigma_{\rm v} \sim$30--450 \kms, significantly lower than that of clusters $\sigma_{\rm v} \sim$ 1000 \kms. This means that morphological transformation effects such as ram-pressure stripping \citep{Gunn_1972} and tidal stripping \citep{Moore_1998}, caused by high velocity encounters with other galaxies or interactions with the IGM, are more likely to occur in dense clusters \citep[e.g.][]{Joshi_2020,Ramatsoku_2020}, but have also been observed in groups \citep[e.g.][]{Sulentic_2001,oosterloo_2018}. Galaxy-galaxy interactions in the form of major and minor mergers are expected to be more likely in the group environment \citep[e.g.][]{Serra_2019} compared to clusters. 
The morphology-density relation \citep{Dressler_1997} captures how `early-type' galaxies (i.e. elliptical and lenticular) are preferably found in high-density environments, such as clusters, and `late-type' galaxies (i.e. spiral and irregular) are found in low-density environments, such as the field, which indicates that a galaxy's environment can be a primary determinant of its gas content and evolution.

Despite the clear correlation between galaxy morphology and environment, several directly related questions remain, including how galaxies evolve in the group environment, specifically their gas content, SFRs and morphologies, and how this differs as a function of group mass. By studying the depletion and removal of cold gas we can understand the transformation of galaxy morphology in dense environments 
\citep[e.g.][]{Sancisi_2008,Schawinski_2014}. Within galaxies, the distribution of neutral hydrogen (\hi) is diffuse and extends well beyond the stellar component of the galaxy \citep[e.g.][]{Broeils_1997,Leroy_2008}. It is therefore a sensitive dynamical tracer and can be used to observe such environmental processes \citep[e.g.][]{Oosterloo_2005,Serra_2012b,Saponara_2017}. 
By observing \hi in galaxies in the group environment, much work has been done to understand the evolution of \hi within this environment \citep[e.g.][]{Yun_1994,Hess_2013,Jones_2019}. Spiral-dominated groups are \hi rich and have a higher galaxy-galaxy interaction rate than elliptical-dominated groups. Elliptical-dominated groups have a higher velocity dispersion (hence, fewer mergers), lower \hi mass, and can have a diffuse, X-ray emitting halo \citep{Freeland_2009}. \hi deficiency in compact groups is linked to a proposed evolutionary sequence where galaxies become increasingly \hi deficient following multiple tidal interactions \citep{Verdes_Montenegro_2001,Borthakur_2015}. The \hi mass function (HIMF) varies significantly with environment \citep[e.g.][]{Rosenberg_2002,Springob_2005,Jones_2020}, and is flatter for galaxy groups at $M_\mathrm{HI} < M^*_\mathrm{HI}$ relative to the global HIMF. This implies a relative deficit of low \hi-mass galaxies and indicates that denser environments, such as groups, are responsible for removing the \hi from low mass galaxies \citep[e.g.][]{Verheijen_2004,Freeland_2009,Busekool_2020}. 
Using an Arecibo Legacy Fast ALFA \citep[ALFALFA; ][]{Giovanelli_2005} sample, consisting of 1613 \hi galaxies in 620 groups in the redshift range of $0.020 < z < 0.042$, sensitivity limited to an \hi mass of $M_\mathrm{HI}\geq10^{8.9}\,$\msun, \cite{Hess_2013} conclude that as the optically-detected membership of a group increases, the \hi-detected galaxies reside increasingly towards the outskirts of the group. As halo mass increases, the fraction of \hi galaxies decreases, and galaxies at the centre of the group are stripped of their gas due to transformation effects. This should be evident in the \hi morphologies, but the \hi detections from ALFALFA are spatially unresolved. In comparison with less dense environments, galaxies located in group-sized dark matter haloes contain $\sim0.4$~dex lower \hi. Haloes with virial masses $M_{\rm vir} >10^{13}\,$\msun have gas fractions of $\log(M_\mathrm{HI}/M_*)\approx-1.3$, while haloes with masses $M_{\rm vir} <10^{13}\,$\msun have gas fractions of $\log(M_\mathrm{HI}/M_*)\approx-0.9$. However, it should be noted that this result is biased to high-mass galaxies with $\log(M_*/\mathrm{M}_\odot)>10$ \citep{Catinella_2013}, and so higher sensitivity surveys, to probe lower \hi masses, will provide a more complete view. 

Even though groups are excellent laboratories for studying the environmental effects on galaxy evolution, they are challenging to identify in 2D images. 
Without reliable spectroscopic redshifts, groups can be difficult to identify and study, because their dark matter haloes are poorly populated \citep{Hess_2013}. This, along with their loose composition make groups difficult to identify via traditional clustering algorithms, especially for low-redshift, low mass systems \citep{Papastergis_2013}. Unlike clusters, their IGM haloes are rarely sufficiently bright to be observed via X-ray emission \citep{Robson_2020} or inverse-Compton scattering via the Sunyaev–Zel'dovich effect \citep{Birkinshaw_1999}, so identification through these standard techniques for clusters is not possible at lower mass, except at a statistical level. Identification and selection via \hi emission at cosmological distances are not readily available as long integration times or very high sensitivities are needed to directly detect the intrinsically faint \hi emission \citep[e.g.][]{Rhee_2017}. The study of \hi in a range of environments requires wide-area surveys at $\sim$few arcsecond angular resolution to spatially resolve galaxies and trace their \hi morphological and dynamical characteristics. This, along with deep \hi surveys, which are key for the statistical study of \hi, will deeply enrich our view of the group assembly and transformation history, particularly for \hi-rich, low-mass galaxy groups. With the sensitivity and frequency coverage of the South African Square Kilometre Array (SKA) precursor, MeerKAT \citep{Jonas_2009}, this will be possible with shorter integration times and over a large redshift range. The MeerKAT International GHz Tiered Extragalactic Exploration (MIGHTEE) Survey \citep{Jarvis_2016} is an ongoing Large Survey Project (LSP) conducted over four deep extragalactic legacy fields, namely COSMOS, XMM-LSS, ECDFS and ELAIS-S1, totalling an area of $\sim$20 deg$^2$. The spectral sensitivity and frequency coverage of this survey allows the detection of \hi in a statistically significant number of galaxies at $z\sim0.2$, as well as rare, massive galaxies out to $z \gtrsim 0.3$. The angular resolution of $\sim 8\,\arcsec$ enables us to study the morphology of individual galaxies at low to intermediate redshifts, adding a powerful diagnostic of environmental effects.

In this paper, we present the discovery of a galaxy group in the XMM-LSS field from the MIGHTEE survey Early Science data. This serendipitous discovery is directly linked to the MIGHTEE-\hi key science goal of studying \hi as a function of environment and demonstrates MeerKAT's suitability for these investigations. This massive group contains 20 \hi galaxies (19 distinct \hi detections) and has not been previously identified despite residing in the well-studied XMM-LSS field. We analyse and discuss the properties of the detected \hi galaxies in the group and compare with their multi-wavelength properties. This paper is organized as follows. In Section~\ref{sec:method} we summarise the specifications of MeerKAT, describe the MIGHTEE observations and summarise the ancillary data. In Section~\ref{sec:results}, we present the total \hi intensity (moment 0) and velocity field (moment 1) maps of the detected \hi sources, a well as their integrated \hi spectra. Using simple models, we also derive the dynamical masses and neutral gas fractions of the \hi sources themselves as well as the group as a whole. Section~\ref{sec:discussion} is a discussion of the morphology and kinematics of the sources and a summary follows in Section~\ref{sec:conclusion}. 

Throughout this work we assume cosmological values of  $\Omega_M=0.315, \Omega_\Lambda=0.685$ and $H_0 = 67.4\,$\kms $\mathrm{Mpc}^{-1}$ \citep{Planck_2020}, resulting in a spatial scale of 866~pc arcsec$^{-1}$ at $z=0.044$.

\section{MeerKAT Observations and Data Reduction}\label{sec:method}

\subsection{MeerKAT}\label{ssec:meerkat}
MeerKAT is a radio interferometer consisting of 64 dishes, located in the Northern Cape, South Africa. Of these 64, a subset of 48 is located within the central core with a diameter of $\sim$1\,km, and the remaining 16 dishes spread out up to maximum baselines of $\sim$8\,km. Each dish has offset-Gregorian optics, consisting of a 13.5\,m diameter main reflector and a 3.8\,m sub-reflector. The receiver indexer is located just below the sub-reflector ensuring a completely unobstructed aperture, which improves the imaging dynamic range by lowering the sensitivity of the primary beam sidelobes to strong sources and radio frequency interference (RFI) outside the main lobe. The large number of MeerKAT antennas, along with the relatively small dish size, results in a high sensitivity array while retaining a large field of view. This makes MeerKAT an excellent facility for wide-field, yet deep imaging surveys. For more detail on the MeerKAT receptors, receivers, primary beam, and correlator see \citet{Camilo_2018,Mauch_2020}.

The MIGHTEE observations in this work use the MeerKAT L-band receiver, which has a usable frequency range from 900 to 1670\,MHz. These Early Science observations use MeerKAT's 4k correlator mode, which has 4096 channels, and a channel width of 209\,kHz, corresponding to 44\kms at $z=0$.

\subsection{The MIGHTEE Survey}\label{ssec:3-mightee}
\begin{figure*}
    \centering
    \includegraphics[scale=0.65]{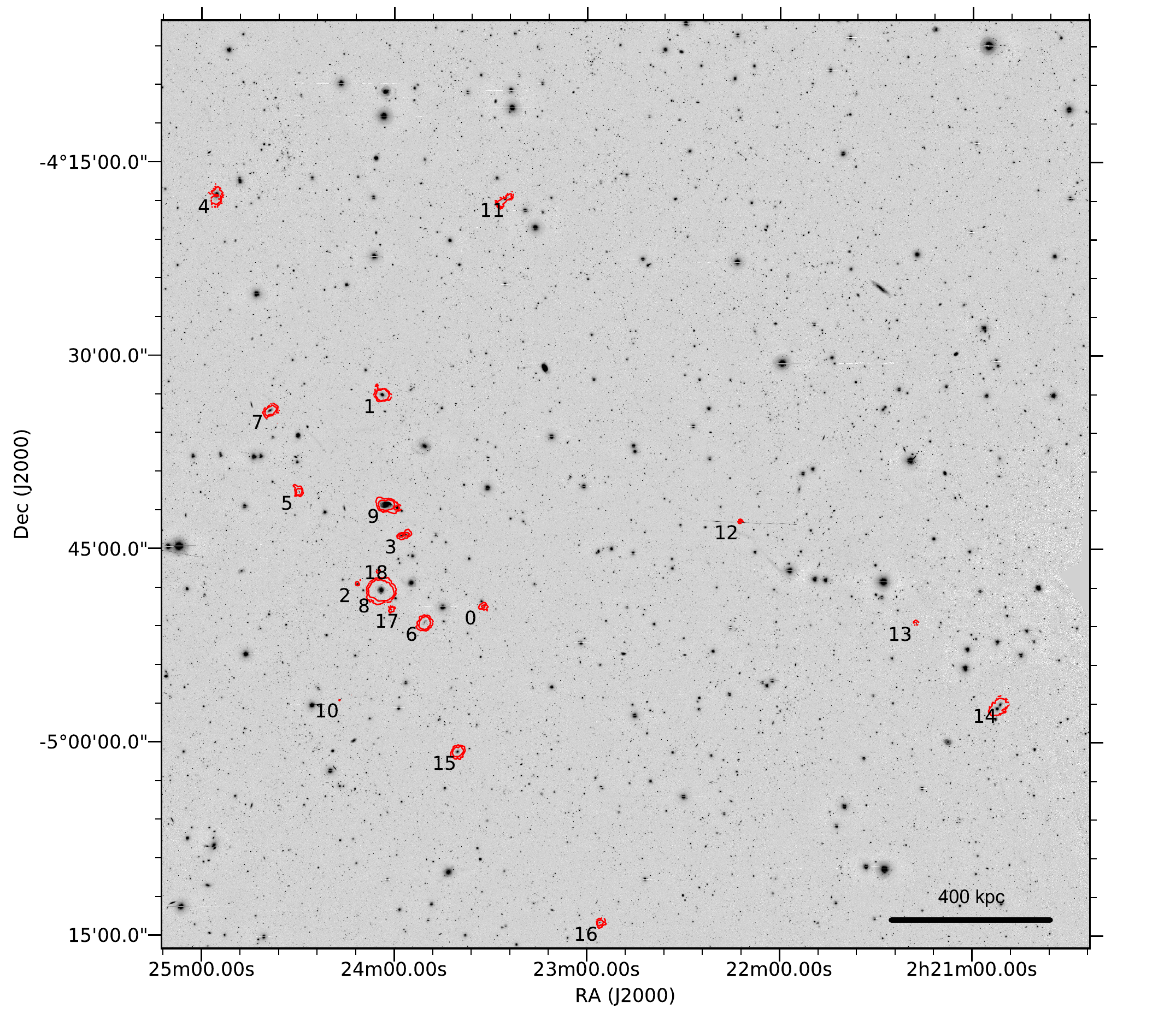}
    \caption[\hi detections in XMM-LSS group: \hi contours and $R$-band]{MeerKAT \hi contours (red) of the 19 detected sources overlaid on the HSC $R$-band image of the XMM-LSS field (greyscale). The \hi contour levels correspond to the column densities of [1.25, 3] $\times 10^{20} \mathrm{cm}^{-2}$. A spatial scale bar is shown in the bottom right.}
    \label{fig:fullimage}
\end{figure*}

%
MIGHTEE \citep{Jarvis_2016} is a deep MeerKAT survey of four legacy multi-wavelength fields. It is one of the LSP's being undertaken by MeerKAT. Given that the MeerKAT correlator has simultaneous continuum and spectral line modes, a significant element within the MIGHTEE survey is an \hi emission survey, referred to as MIGHTEE-HI. MeerKAT's high sensitivity and wide field-of-view enable deep interferometric \hi surveys over a significant number of extragalactic legacy fields. For more detail on the MIGHTEE-HI component of the survey, see \citet{Maddox_2021} and Frank et al. (in prep.).
Within Early Science, a galaxy group at $z\sim 0.044$ was discovered in the XMM-LSS field. The \hi-detected galaxies in the group are shown in Fig.~\ref{fig:fullimage}, and the details of the observation are summarised in Table~\ref{tab:observation}.

\begin{table}
\centering
\caption{Summary of MeerKAT MIGHTEE data products used in this work (i.e. limited to the study of this galaxy group). More detail on the observations themselves, as well as the data processing is described in Frank et al.~(in prep.).}
\begin{tabular}{ll}
\hline
\hline
Channel width                            & 209 kHz                                                    \\
                                         & 46 \kms for \hi at z = 0.044 \\
Pixel size                       & 2 arcsec                                                   \\
\hi cube weighting        & Briggs \textsc{robust} = 0.5              \\
Median \hi channel rms noise & 85 $\mu$Jy \perbeam                         \\
Restoring beam FWHM                      & 12 $\times$ 9 arcsec$^2$                                    \\
                                         & PA = -30 deg                                               \\
On-source integration time & 12 hours \\
$N_\mathrm{HI}$ sensitivity ($3\sigma$)  & 1.6 $\times \, 10^{20}\, \mathrm{cm}^{-2}$ (per channel)                \\ \hline
\end{tabular}
\label{tab:observation}
\end{table}

\subsection{Calibration and Imaging}\label{ssec:calibration}
%
MIGHTEE-HI data products were produced using the {\sc ProcessMeerKAT} calibration and imaging pipeline, described in detail in Frank et al.~(in prep.). Very briefly, this is a parallelised {\sc Casa}-based\footnote{http://casa.nrao.edu} pipeline run within {\sc Singularity}\footnote{https://singularity.lbl.gov} containers on the ILIFU Cloud facility\footnote{http://www.ilifu.ac.za}. The pipeline does the continuum and polarimetric calibration, following standard calibration routines and strategies, including flagging, delay, bandpass, and complex gain calibration, with automated self-calibration in development. The continuum cross-calibration was done at a lower frequency resolution (averaging by a factor of 4), and was followed by self-calibration. The resultant cross and self-calibration gains were applied to the full resolution data, followed by a final bandpass calibration. 

Wide-field, spectral-line imaging was performed using {\sc Casa}'s \texttt{TCLEAN} task with a Briggs ({\sc robust=0.5}) weighting scheme. Continuum subtraction was done in the visibility and the image domain. The continuum model was subtracted using \texttt{UVSUB} and was followed by subtracting a fit to the residual visibilities using \texttt{UVCONTSUB}. Once the cubes were produced, per-pixel median filtering was implemented on the cubes to reduce the impact of direction-dependent artefacts. The final, continuum-subtracted \hi spectral line sub-cube has an angular extent of 2.3$\times$2.3~deg$^2$ and spans 1330.198 -- 1379.936 MHz. The average Point Spread Function FWHM for the cubes is 9$\times$12\,arcsec, corresponding to a spatial scale of 7.8$\times$10.4\,kpc for the galaxy group members at $z=0.044$. 
%
\subsection{Ancillary Data}\label{ssec:3-ancdata}
%


One of the primary motivations for targeting the legacy fields is the wealth of multi-wavelength measurements and derived properties that are available for a large number of galaxies. Amongst the primary data products in the XMM-LSS field are from the Hyper SuprimeCam Subaru Strategic Program \citep[HSC-SSP,][]{Aihara_2017} and Canada-France-Hawaii Telescope Legacy Survey \citep[CFHTLS; ][]{Cuillandre_2012} {in optical wavelengths}, and the VISTA VIDEO \citep{Jarvis_2012} survey in the near-infrared. A detailed description of these is provided in \citet{Maddox_2021}.

{The MIGHTEE-HI team extracted fluxes in the $ugrizYJHK_{s}$-bands from the HSC and VIDEO images (image reduction is described in \citealt{Bowler_2020}). Together, this provides a wavelength coverage of $\sim0.4\text{-}2\,\mu$m. Since our sources are highly extended and relatively few in number, we manually fit elliptical apertures to each source, from which flux is extracted. The SED-fitting code LePhare \citep{Arnouts_1999,Ilbert_2006} was used to determine the stellar masses and star-formation rates for each of the detected objects following the same process used in \citet{Adams_2021}. To summarise, we begin by fixing the redshifts of each source to those determined from the HI line $z_{\rm HI}$ (see Section~\ref{ssec:3-hidetections} and Table~\ref{tab:results}). We then fit synthetic templates of galaxy spectra derived from \citet{Bruzual_2003} using minimisation of $\chi^2$ with LePhare. These synthetic spectra are generated with star formation histories following either a constant star formation rate or exponentially decaying star formation rates with timescales $\tau = [0.1, 0.3, 1, 2, 3, 5, 10, 15, 30]$ Gyr. These template spectra are further modified by the effects of redshifting and various levels of dust extinction \citep{Calzetti_2000}. The result of this process provides the best fitting galaxy SED along with the associated stellar mass and star formation rate for that template (Table~\ref{tab:results}).}
%

\section{Results}\label{sec:results}

\subsection{HI Detections}\label{ssec:3-hidetections}
Source finding was done visually and unguided by members of the MIGHTEE-HI Working Group (WG), as described in \citet{Maddox_2021}. We find 19 \hi detections in the region we associate with the galaxy group, both spatially and spectrally. The 19 detections included 20 galaxies, two of which are a closely interacting pair. 
All of the detected \hi sources are shown in Fig.~\ref{fig:fullimage}, where the $1.25 \times 10^{20}\,\mathrm{cm}^{-2}$ column density contour is overlaid on the $R$-band HSC image. The galaxies are labelled as listed in Table~\ref{tab:results}. The moment 0 maps, moment 1 maps, and the integrated \hi spectra of these sources are shown in Fig.~\ref{fig:3-mommap1}. The RGB images in these figures are comprised of the HSC $G$, $R$, and $I$ bands. The sources reside in the frequency range of 1330.198 -- 1379.936\,MHz, which corresponds to a recession velocity range of 12,679 -- 13,093\kms, and a redshift range of 0.043 -- 0.045. All \hi detections have a peak channel flux density signal-to-noise ratio above $4\sigma$, and are optically visible, with aperture-defined $R$-band magnitudes ranging from 20.0 > mag$_r$ > 13.9. The moment 0 maps were constructed according to the following procedure.
\begin{enumerate}
    \item Each cubelet centred on a detection was smoothed to a circular beam of 20" and clipped at $3\sigma$ level.
    \item The resulting mask was applied to the original resolution cubelet to account for the low column density extended \hi emission.
    \item Every moment 0 map was examined by eye and isolated by masking out the noise peaks if any were present within the field of view.
\end{enumerate}
The spectra were extracted from the unsmoothed cube in the regions defined by the mask. 
We note that the continuum subtraction might result in a slight suppression of the integrated flux around the emission region of large sources, this effect was not accounted for in our work.
%
\begin{figure*}
    \centering            
    \includegraphics[width=0.75\textwidth]{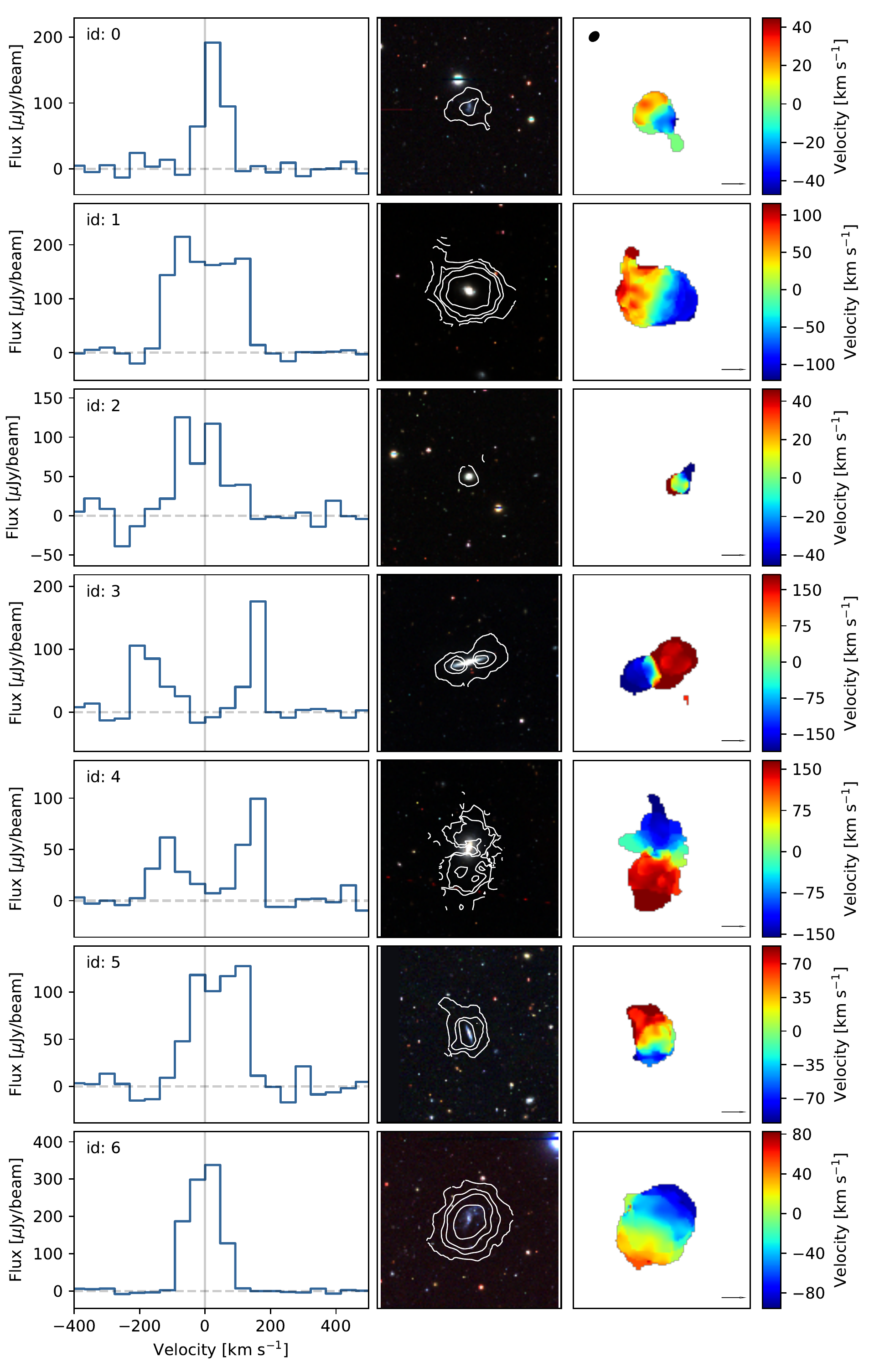}
    \caption[Moment maps and spectra]{From left to right, the spectra, moment 0 contours plotted over RGB cutouts from the HSC survey, and moment 1 maps of source ID's 0--6 of the 19 detected \hi sources in the group. The Gaussian restoring beam in the top left (black) has a mean FWHM of $12 \times 9 \,\mathrm{arcsec}^2$ and the spatial scale bar in the bottom right indicates 20 kpc. The contours shown correspond to \hi column densities of [1, 3, 5, 10] $\times 10^{20} \mathrm{cm}^{-2}$.}
    \label{fig:3-mommap1}
\end{figure*}
\renewcommand{\thefigure}{\arabic{figure} (cont.)}
\begin{figure*}
    \ContinuedFloat
    \captionsetup{list=off,format=cont}
    \centering
    \includegraphics[width=0.75\textwidth]{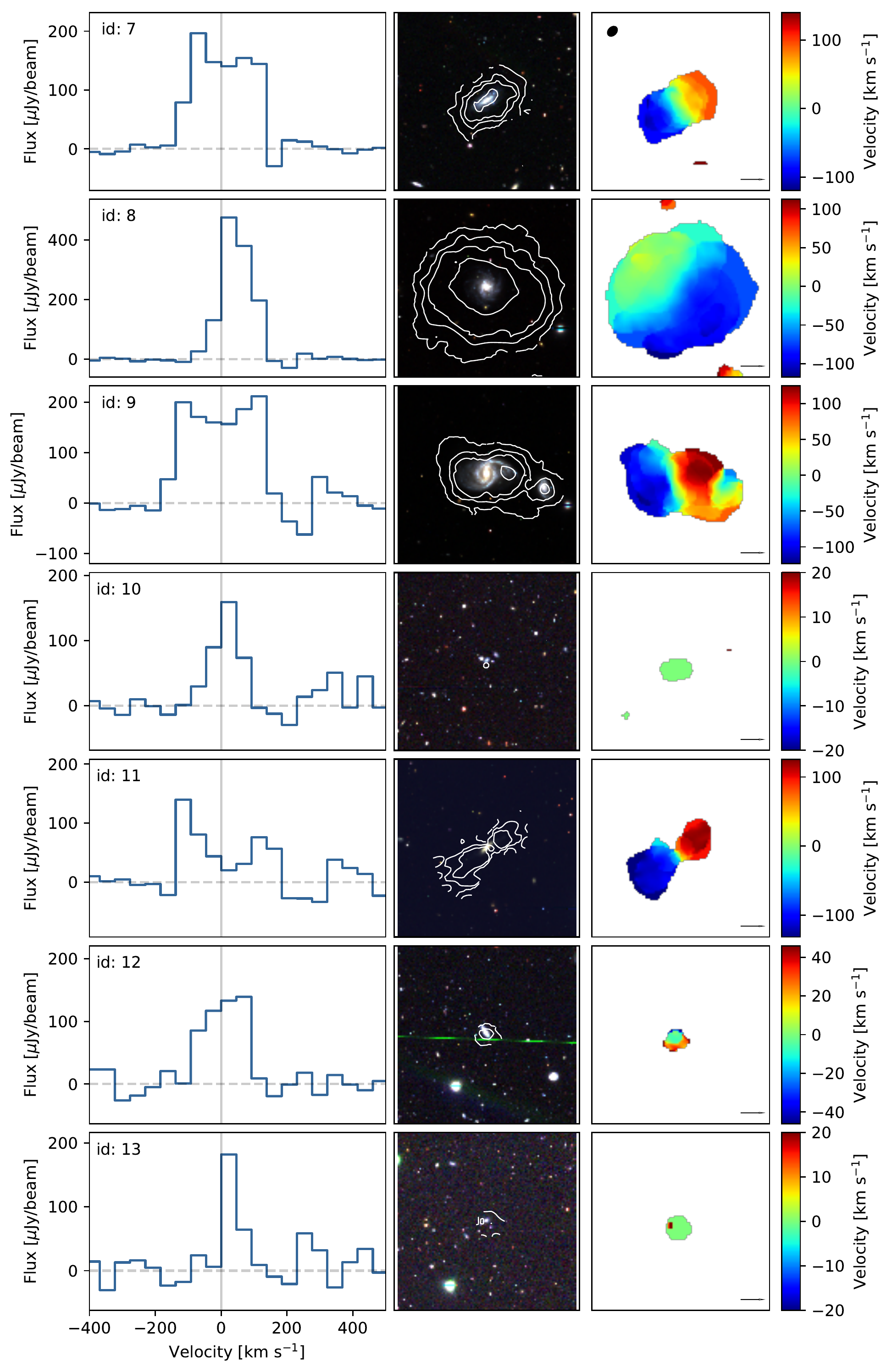}
    \caption[Moment maps and spectra]{As per Fig.~\ref{fig:3-mommap1}, but for source ID's 7--13.}
    \label{fig:3-mommap2}
\end{figure*}

\begin{figure*}
    \ContinuedFloat
    \captionsetup{list=off,format=cont}
    \centering
    \includegraphics[width=0.75\textwidth]{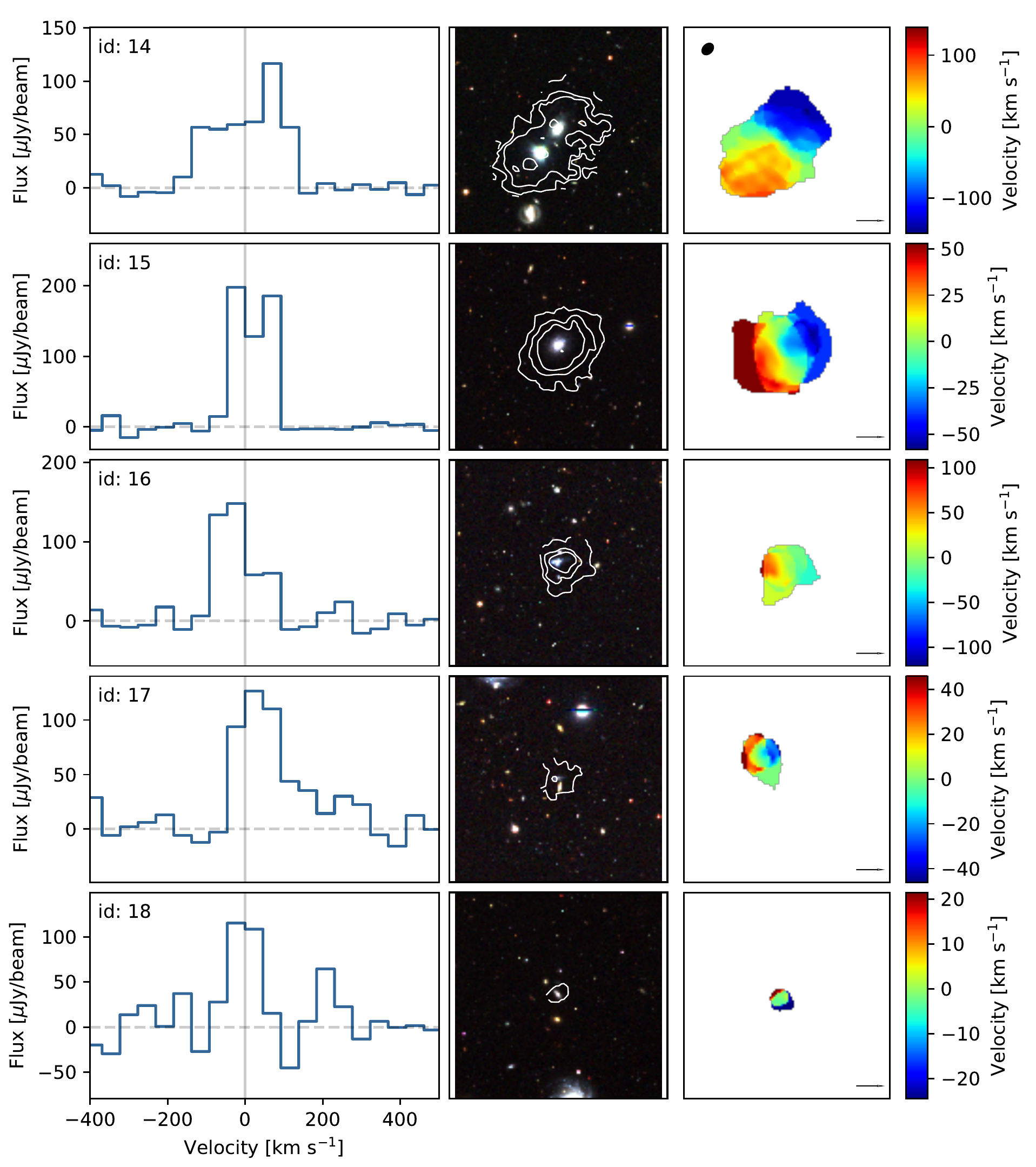}
    \caption[Moment maps and spectra]{As per Fig.~\ref{fig:3-mommap1}, but for source ID's 14--18.}
    \label{fig:3-mommap3}
\end{figure*}

\renewcommand{\thefigure}{\arabic{figure}}
%

The stellar masses $M_*$ and SFR's of the detected sources were measured from the SED fitting of the multi-wavelength photometry, as discussed in Section~\ref{ssec:3-ancdata}. The stellar masses have a conservative uncertainty estimate of $\pm$0.1 dex. The objects were visually cross-matched with spatial coordinates by the MIGHTEE-HI WG.
\begin{table*}
\centering 
\caption[Catalogue of \hi detected sources in group]{Catalogue of \hi-detected sources.}
\resizebox{\textwidth}{!}{\begin{tabular}{lllllrrrrrrrrr} 
\hline ID	&	RA (J2000)	&	Dec (J2000)	&	$z_\mathrm{HI}$	&	{	$D_L$	}	&	log			&	{	$w_{50}$	}	&	$d_\mathrm{HI}$	&	{	i	}	&	log	&	{	log(SFR/	}	&	$f_\mathrm{HI}^*$	&	log	&	$f_\mathrm{HI}^\mathrm{dyn}$	\\
	&	h:m:s	&		&		&	{	[Mpc]	}	&	($M_\mathrm{HI}/\mathrm{M}_\odot$)			&	{	[km s$^{-1}$]	}	&	[kpc]	&	{ [deg]	}	&	($M_*/\mathrm{M}_\odot$)	&	{	M$_\odot$ yr$^{-1}$)	}	&		&	($M_\mathrm{dyn}/\mathrm{M}_\odot$)	&		\\
(1)	&	(2)	&	(3)	&	(4)	&	{	(5)	}	&	(6)			&	{	(7)	}	&	(8)	&	{	(9)	}	&	(10)	&	{	(11)	}	&	(12)	&	(13)	&	(14)	\\ \hline
4	&	02:24:55.35	&	-04:17:35.25	&	0.0440	&	{	201.4	}	&	10.076	$\pm$	0.004	&	{	462	}	&	91	&	{	39.41	}	&	10.76	&	{	-0.16	}	&	-0.68	&	11.81	&	-1.73	\\
7	&	02:24:38.69	&	-04:34:18.06	&	0.0437	&	{	199.8	}	&	9.905	$\pm$	0.002	&	{	225	}	&	51	&	{	67.70	}	&	9.47	&	{	-0.13	}	&	0.43	&	10.99	&	-1.08	\\
5	&	02:24:29.62	&	-04:40:36.72	&	0.0440	&	{	201.4	}	&	9.504	$\pm$	0.005	&	{	181	}	&	19	&	{	78.76	}	&	8.55	&	{	-0.84	}	&	0.95	&	10.39	&	-0.89	\\
10	&	02:24:17.02	&	-04:56:42.19	&	0.0436	&	{	199.1	}	&	8.476	$\pm$	0.036	&	{	41	}	&	-	&	{	42.65	}	&	7.71	&	{	-0.58	}	&	0.77	&	-	&	-	\\
2	&	02:24:11.49	&	-04:47:44.84	&	0.0445	&	{	203.6	}	&	8.600	$\pm$	0.033	&	{	124	}	&	10	&	{	50.64	}	&	9.42	&	{	-0.11	}	&	-0.82	&	9.85	&	-1.25	\\
18	&	02:24:04.97	&	-04:46:47.45	&	0.0442	&	{	202.1	}	&	8.421	$\pm$	0.045	&	{	201	}	&	11	&	{	56.65	}	&	8.85	&	{	-0.90	}	&	-0.43	&	10.22	&	-1.80	\\
8	&	02:24:04.13	&	-04:48:12.50	&	0.0436	&	{	199.1	}	&	10.664	$\pm$	0.000	&	{	248	}	&	115	&	{	32.76	}	&	10.28	&	{	0.71	}	&	0.38	&	11.41	&	-0.75	\\
1	&	02:24:03.77	&	-04:33:05.07	&	0.0442	&	{	202.1	}	&	10.221	$\pm$	0.001	&	{	276	}	&	63	&	{	54.51	}	&	10.18	&	{	0.73	}	&	0.04	&	11.24	&	-1.02	\\
9a	&	02:24:02.64	&	-04:41:36.30	&	0.0434	&	{	198.3	}	&	10.089	$\pm$	0.001	&	{	66	}	&	74	&	{	40.37	}	&	11.00	&	{	0.57	}	&	-0.84	&	10.30	&	-0.14	\\
17	&	02:24:00.94	&	-04:49:43.42	&	0.0440	&	{	201.4	}	&	8.951	$\pm$	0.013	&	{	218	}	&	-	&	{	70.21	}	&	8.71	&	{	0.25	}	&	0.24	&	-	&	-	\\
9b	&	02:23:59.20	&	-04:41:48.15	&	0.0434	&	{	198.3	}	&	9.344	$\pm$	0.004	&	{	380	}	&	33	&	{	42.81	}	&	10.20	&	{	0.22	}	&	-0.86	&	-	&	-	\\
3	&	02:23:57.07	&	-04:43:57.22	&	0.0439	&	{	200.6	}	&	9.494	$\pm$	0.004	&	{	15	}	&	-	&	{	78.39	}	&	10.65	&	{	0.32	}	&	-1.15	&	-	&	-	\\
6	&	02:23:50.63	&	-04:50:45.02	&	0.0440	&	{	201.4	}	&	10.117	$\pm$	0.001	&	{	98	}	&	52	&	{	71.54	}	&	8.57	&	{	-0.53	}	&	1.55	&	10.40	&	-0.28	\\
15	&	02:23:40.35	&	-05:00:45.18	&	0.0447	&	{	204.4	}	&	9.917	$\pm$	0.001	&	{	157	}	&	57	&	{	41.93	}	&	9.47	&	{	0.03	}	&	0.45	&	10.77	&	-0.85	\\
0	&	02:23:32.21	&	-04:49:27.54	&	0.0447	&	{	204.4	}	&	9.191	$\pm$	0.007	&	{	101	}	&	25	&	{	43.70	}	&	8.70	&	{	-1.27	}	&	0.49	&	10.11	&	-0.92	\\
11	&	02:23:25.54	&	-04:17:52.68	&	0.0432	&	{	197.6	}	&	9.801	$\pm$	0.006	&	{	243	}	&	-	&	{	80.77	}	&	10.21	&	{	0.07	}	&	-0.40	&	-	&	-	\\
16	&	02:22:56.00	&	-05:13:59.47	&	0.0447	&	{	204.4	}	&	9.452	$\pm$	0.006	&	{	85	}	&	30	&	{	62.64	}	&	7.96	&	{	0.96	}	&	1.49	&	10.06	&	-0.61	\\
12	&	02:22:12.31	&	-04:42:51.79	&	0.0442	&	{	202.1	}	&	8.716	$\pm$	0.026	&	{	154	}	&	11	&	{	60.33	}	&	9.09	&	{	-0.57	}	&	-0.38	&	10.05	&	-1.33	\\
13	&	02:21:17.84	&	-04:50:40.48	&	0.0436	&	{	199.1	}	&	8.726	$\pm$	0.041	&	{	58	}	&	11	&	{	52.19	}	&	7.71	&	{	-1.85	}	&	1.02	&	9.41	&	-0.68	\\
14	&	02:20:51.46	&	-04:57:03.72	&	0.0440	&	{	201.4	}	&	10.279	$\pm$	0.002	&	{	331	}	&	97	&	{	44.59	}	&	9.92	&	{	-0.18	}	&	0.36	&	11.57	&	-1.29	\\ \hline 
\multicolumn{14}{l}{\footnotesize  Note: Cols (2)--(3): RA and Dec (J2000) coordinates, based on the optical counterparts of the \hi detections. Cols (4)--(5): \hi redshifts and luminosity distances. Cols (6)--(9)}\\
\multicolumn{14}{l}{\footnotesize  \hi mass, corrected FWHM of integrated \hi line profile, \hi disk diameter, and inclination, as described in Section~\ref{ssec:3-hidetections}. Cols (10)--(11): Stellar mass and star-formation rates}\\
\multicolumn{14}{l}{\footnotesize determined from SED fitting. Col (12): \hi-to-stellar mass ratio. Col (13): Dynamical mass. Col (14): \hi-to-dynamical mass ratio. Uncertainties on d, $M_\mathrm{HI}$ and W$_{50}$ are}\\
\multicolumn{14}{l}{\footnotesize discussed in Section~\ref{sec:results}.}\\
\end{tabular}}
\label{tab:results}
\end{table*}
%

 The \hi redshift $z_\mathrm{HI}$ was extracted from the central channel of the emission profile of each source, and the \hi mass was calculated according to \citet{Meyer_2017}:
\begin{equation}
\left(\frac{M_{\mathrm{HI}}}{M_{\odot}}\right) \simeq \frac{2.356 \times 10^{5}}{(1+z)}\left(\frac{D_{L}}{\mathrm{Mpc}}\right)^{2}\left(\frac{S}{\mathrm{Jy}\,\mathrm{km\,s}^{-1}}\right)
\end{equation}
where $D_L$ is the cosmological luminosity distance to the source, and $S$ is the integrated \hi flux density, determined from the moment-0 maps. The luminosity distance was calculated for each source, based on its \hi redshift, and assuming the cosmological values stated in Section~\ref{sec:intro}. At the distance of these sources ($\sim 200$ Mpc), peculiar motions are negligible compared to the Hubble flow, so the redshift can be converted directly to a distance. The error on the integrated flux $S$ was measured by applying the 3D source mask to the four emission-free regions around the detection and by calculating the mean rms (Table~\ref{tab:results}).  Inclination angles of each source were defined as $\cos^2(i)=(b^2-\theta_{b}^2)/(a^2-\theta_{a}^2)$, where $a$ and $b$ are the major and minor axis of an ellipse, fitted to the outer contour of the \hi moment 0 map, $\theta_{a}$ and $\theta_{b}$ are the sizes of the synthesised beam in the direction of the major and minor axis of the \hi disk, used to correct for the beam-smearing effect \citep{VerheijenSancisi2001}. For simplicity we assume an infinitely thin \hi disk. The \hi diameter ($d_{\rm HI}$) of each galaxy was measured by fitting an ellipse to the contour of the inclination-corrected \hi column density map corresponding to $\rm 1 [M_{\odot} \, pc^{-2}] = 1.249 \times 10^{20} [cm^{-2}]$ \citep{Ponomareva16}. A half beam size of 5 kpc was assigned as an uncertainty of the d$_{\rm HI}$.

The full-width half maximum of the integrated \hi line profile ($\rm W_{50}$) of each galaxy was attained by fitting Busy Functions (BF) \citep{Westmeier_2013} using \textsc{PyMultiNest} \citep{Buchner_2014}, a Monte Carlo library that uses Bayesian analysis for parameter estimation. Considering that the Early Science data has coarse channel width, a bootstrap technique was implemented to solve for and fix the BF parameters during different \textsc{PyMultiNest} runs. The maximum likelihood method was then used to identify the best-fit model (Mulaudzi et al.~in prep.). 
{The resulting $\rm W_{50}$ values were corrected for projection effects ($\mathrm{W_{50}}/\sin(i)$), instrumental broadening and turbulent motions following the standard prescriptions \citep{VerheijenSancisi2001}. Specifically, the instrumental broadening correction for our data is $\sim 26$ \kms, while $5$ \kms was adopted as a standard correction for the turbulent motions when $\rm W_{50}$ is matched to $\rm V_{max}$ \citep{Ponomareva16}.}
The resulting measurements are summarised in Table~\ref{tab:results}. The channel width was assigned as the typical error on the $\rm W_{50}$ measurement. 


\subsection{Properties of individual {sources}}
Given the coarse frequency resolution of these Early Science data, combined with the relatively few spatial resolution elements across each source, we restricted ourselves to a simple dynamical mass modelling approach. 
Following \citet{de_Blok_2014}, the dynamical mass is estimated using
\begin{equation}
\label{eq:dynanicalmass}
M_\mathrm{dyn} = \frac{R}{\mathrm{G}}\left(\frac{\rm W_{50}}{2 \sin i }\right) ^2,
\end{equation}
where $R = d_\mathrm{HI}/2$, and W$_{50}$ and $i$ are determined as described in Section~\ref{ssec:3-hidetections}. These dynamical masses are listed in Table~\ref{tab:results} and have typical uncertainties of $^{+0.14}_{-0.19}$dex, which are merely intended to be indicative and are estimated by channel velocity width and the positional uncertainty related to the FWHM of the restoring beam.

\subsection{Properties of the entire galaxy group}
 The non-Gaussian velocity distribution of the HI-selected galaxies show a group that is clearly not yet virialised (see Fig.~\ref{fig:hist}). i.e. the group is still in the process of assembly. However, the close proximity in space and redshift suggest that the vast majority, if not all, of these galaxies are bona fide group members. For brevity, we refer to all galaxies presented here as group members, mindful of the caveat that a handful may possibly still be in the process of joining the system.

For all galaxies, we calculate the \hi-to-stellar mass fraction, which is defined as $f_\mathrm{HI}^*\equiv \log (M_\mathrm{HI}/M_*)$. We also calculate the \hi-to-dynamical mass fraction, which is defined as $f_\mathrm{HI}^\mathrm{dyn}\equiv \log(M_\mathrm{HI}/M_\mathrm{dyn})$. These results are displayed in Table~\ref{tab:results}. The total \hi mass in the group is $\log(M_\mathrm{HI}/\mathrm{M}_\odot) = 11.20 \pm 0.01$, assuming all \hi mass is detected. By summing the dynamical masses of the individual galaxies, the total dynamical mass of galaxies in the group is $\log(M_\mathrm{dyn,gal}^\mathrm{HI}/\mathrm{M}_\odot) = 12.32 \pm 0.23$, typical for an intermediate-mass galaxy group \citep{Tully_2015}, assuming that the group is solely comprised of the galaxies detected in \hi. From this, the total \hi gas mass fraction of galaxies in the group is $f_\mathrm{HI,tot}^\mathrm{dyn}=-1.26$. 
The ratio between the sum of the stellar mass and \hi mass to dynamical mass is $f_\mathrm{HI+*,tot}^\mathrm{dyn}=-0.90$ (12.6\%), which is broadly consistent with the baryon fraction of intermediate halo mass groups, that is $\sim 13.6 \%$ for $M_\mathrm{halo,vir} = 2.1\times 10^{13} \,\mathrm{M}_\odot$ \citep{Giodini_2009}. 

\begin{figure*}
    \centering            
    \includegraphics[scale=0.6]{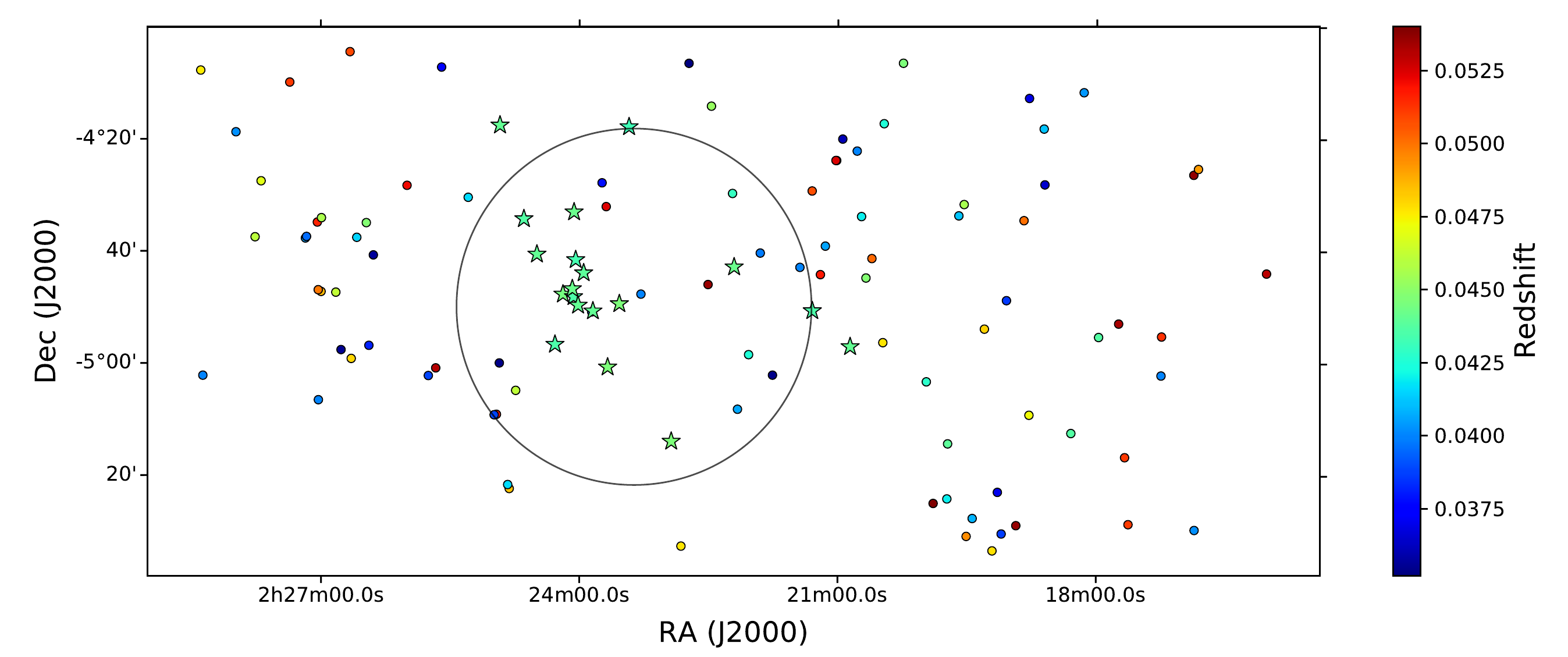}
    \caption{The spatial and redshift distribution of other known nearby sources in the XMM-LSS field. There are 92 sources in total, of which 13 fall within the beam FWHM of this pointing. The black circle shows the FWHM of the primary beam. The stars represent the \hi-detected group members. The colourbar shows the photometric redshift of the non-detected sources (\hi redshift for detected sources), which have large uncertainties $\Delta z = 0.02$. Their group membership is therefore uncertain.}
    \label{fig:other}
\end{figure*}

In the above discussion, we assume that the \hi galaxies comprise all of the galaxies in the group. However, \hi-deficient galaxies obviously may be missing with this selection. We therefore use the deep optical and near-infrared imaging, along with photometric redshifts to determine whether there may be a large number of \hi-deficient group members. Fig.~\ref{fig:other} shows the potential galaxy group members based on their photometric redshift probability density functions \citep{Bowler_2020,Adams_2020,Adams_2021}, where we assume that a galaxy is a potential group member if its photometric redshift lies in the redshift range $0.035<z<0.055$ and $g-$band magnitude $<21.5$. Within the primary beam of the MIGHTEE observations discussed here, we find 13 galaxies which meet this criteria.  The photo-z uncertainty ($\Delta z = 0.02$) is too large to associate, with high probability, any of these optically-selected sources with the HI galaxy group. Although we cannot rule out any of these galaxies being members of the group, the density of such sources is similar to that across the wider XMM-LSS field, as such we do not find strong evidence for a large number of \hi-deficient galaxies in the group.  The availability of spectroscopic redshifts for HI-deficient sources in the vicinity of the group will provide a better assessment of group membership.

\section{Discussion}\label{sec:discussion}
%
The majority of the identified \hi sources in Fig.~\ref{fig:3-mommap1} can be classified as spiral or disk-like galaxies, based on the optical images, with the exception of sources ID 0 and ID 6, which appear to be irregular galaxies. Unlike the other sources, sources ID 0 and ID 6 do not have a sign of spiral structure in the optical images. Sources ID's 2, 10, 13, 17 and 18 have amongst the lowest \hi masses of the detected sources ($8.4 < \log(M_\mathrm{HI}/\mathrm{M}_\odot) < 8.8$). The optical images do not show structure within these galaxies. 
Object ID's 1, 2, 3, 4, 5, 7, 9, 11, and 15 exhibit the characteristic double-horned spectral profile of an inclined spiral galaxy, with $\rm W_{50}$ parameters that range between 119--313 \kms. 

\begin{figure}
\centering
\includegraphics[width=0.9\columnwidth]{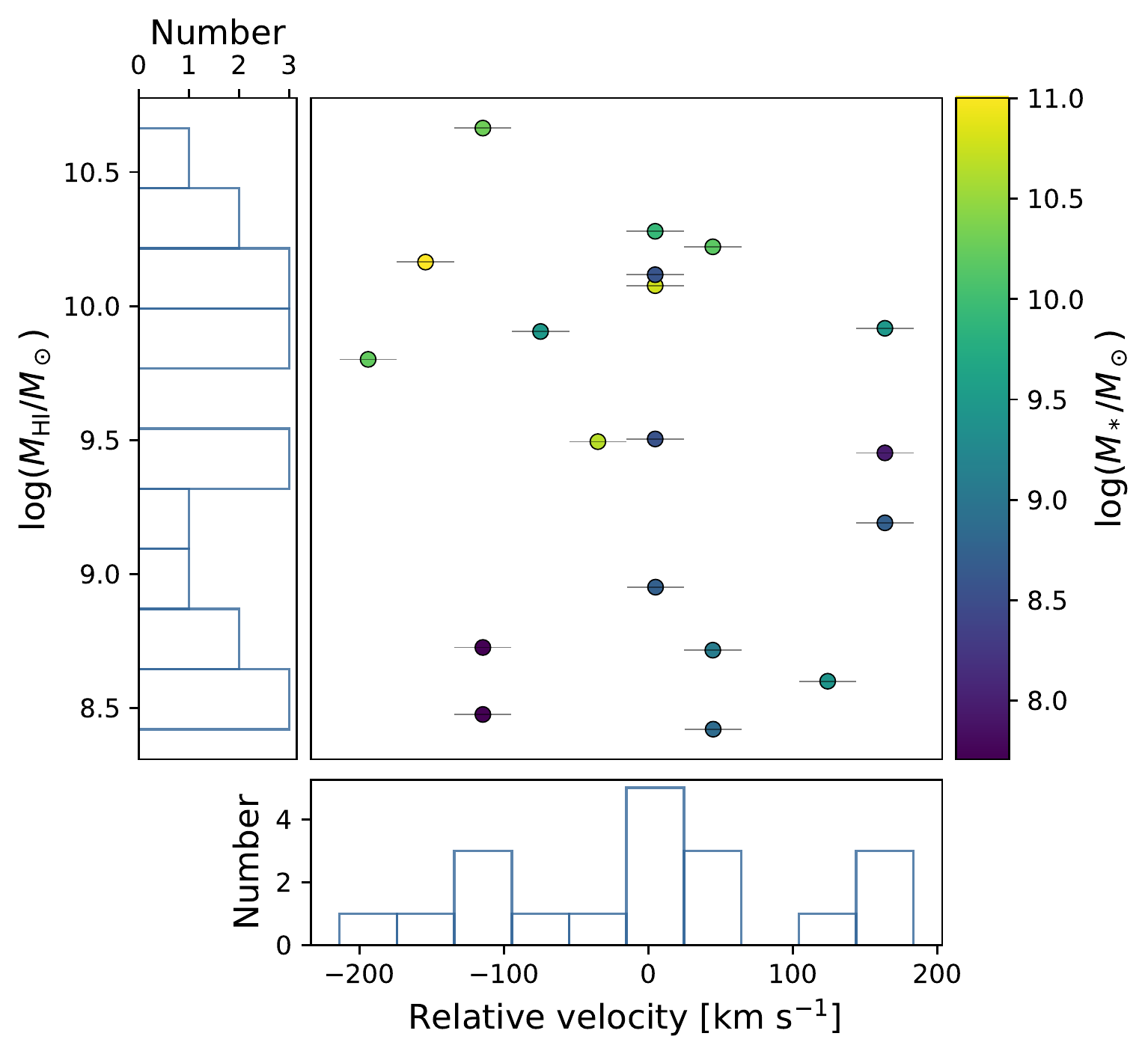}
\caption[Histogram of \hi masses and redshifts for galaxies in group]{Histograms showing the distribution \hi masses and velocities relative to the group centre of the 19 detected sources. The colour of the points corresponds to the stellar mass.}
\label{fig:hist}
\end{figure}
In most cases, the \hi extent of the galaxy is significantly larger than the optical. Most notably, sources ID 6 and ID 8 with \hi radii approximately two and four times that of their $R$-band radii respectively. Source ID 1 is faint and diffuse in the optical, with clumpy star-formation in the arms, but with high column density \hi ($N_\mathrm{HI} \leq 2.5 \times 10^{21} \,\mathrm{cm}^{-2}$). Clumpy star-formation regions can also be seen in sources ID 7 and ID 8. All detected galaxies have a typical \hi mass of $ 8 < \log(M_\mathrm{HI}/\mathrm{M}_\odot) < 11$ (as illustrated in Fig. \ref{fig:hist}). The higher-mass outlier, source ID 6, has an \hi mass of $\log(M_\mathrm{HI}/\mathrm{M}_\odot) = 10.66$, which is on the high-mass end of the \hi-mass function $\log(M_\mathrm{HI}^*/\mathrm{M}_\odot) > 9.94 \pm 0.02 \pm 0.05 h_{70}^{-2}$ \citep{Jones_2018}. This source also has the highest \hi-to-stellar mass ratio of the group of $f_\mathrm{HI}^* = 1.55$, which is interesting, as the most \hi-massive objects tend to be stellar-dominated \citep{Maddox_2014}.

\begin{figure*}
\centering
\subfloat{%
\centering
  \includegraphics[width=0.7\columnwidth]{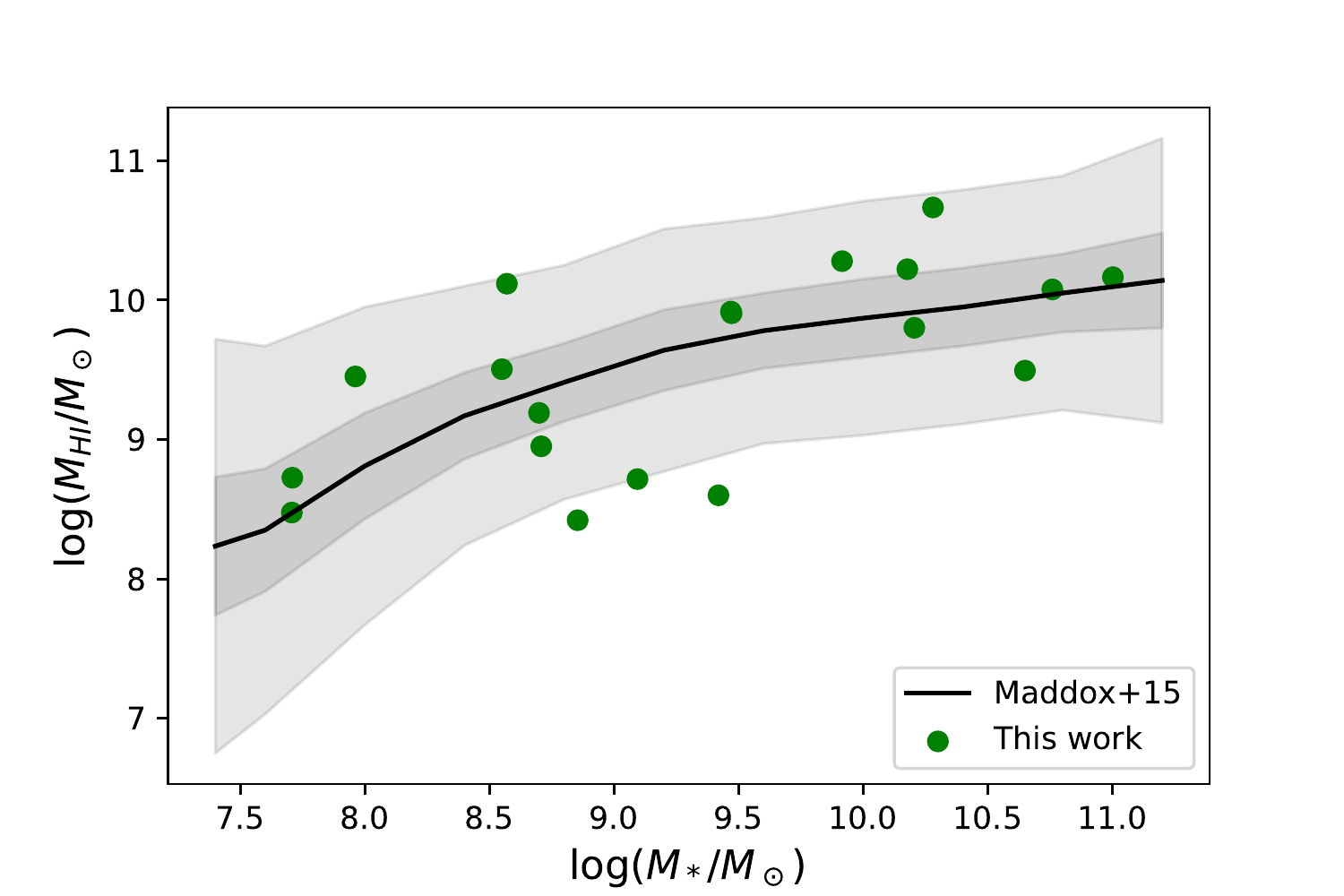}%
}
\subfloat{%
\centering
  \includegraphics[width=0.7\columnwidth]{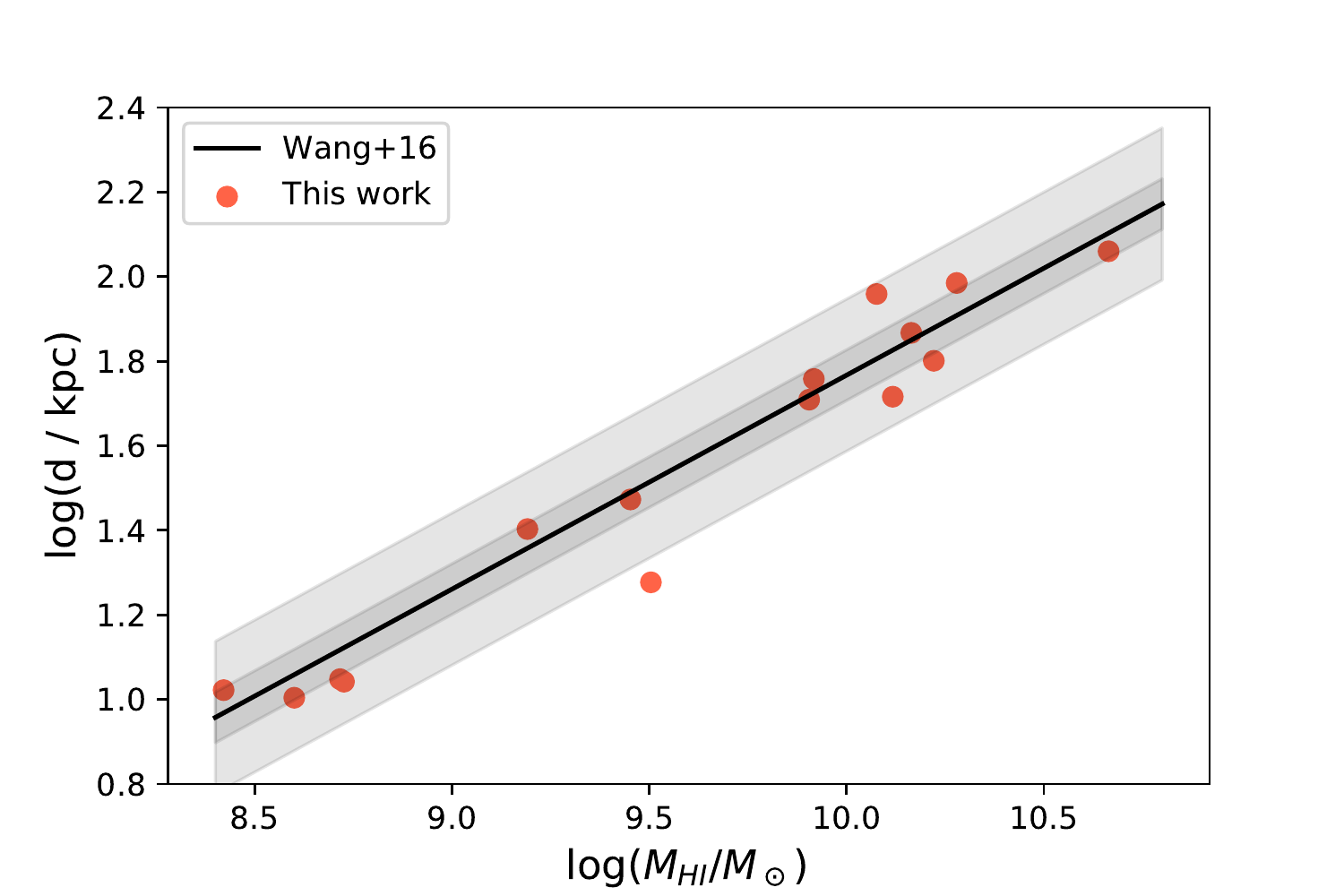}%
}
\subfloat{%
\centering
  \includegraphics[width=0.7\columnwidth]{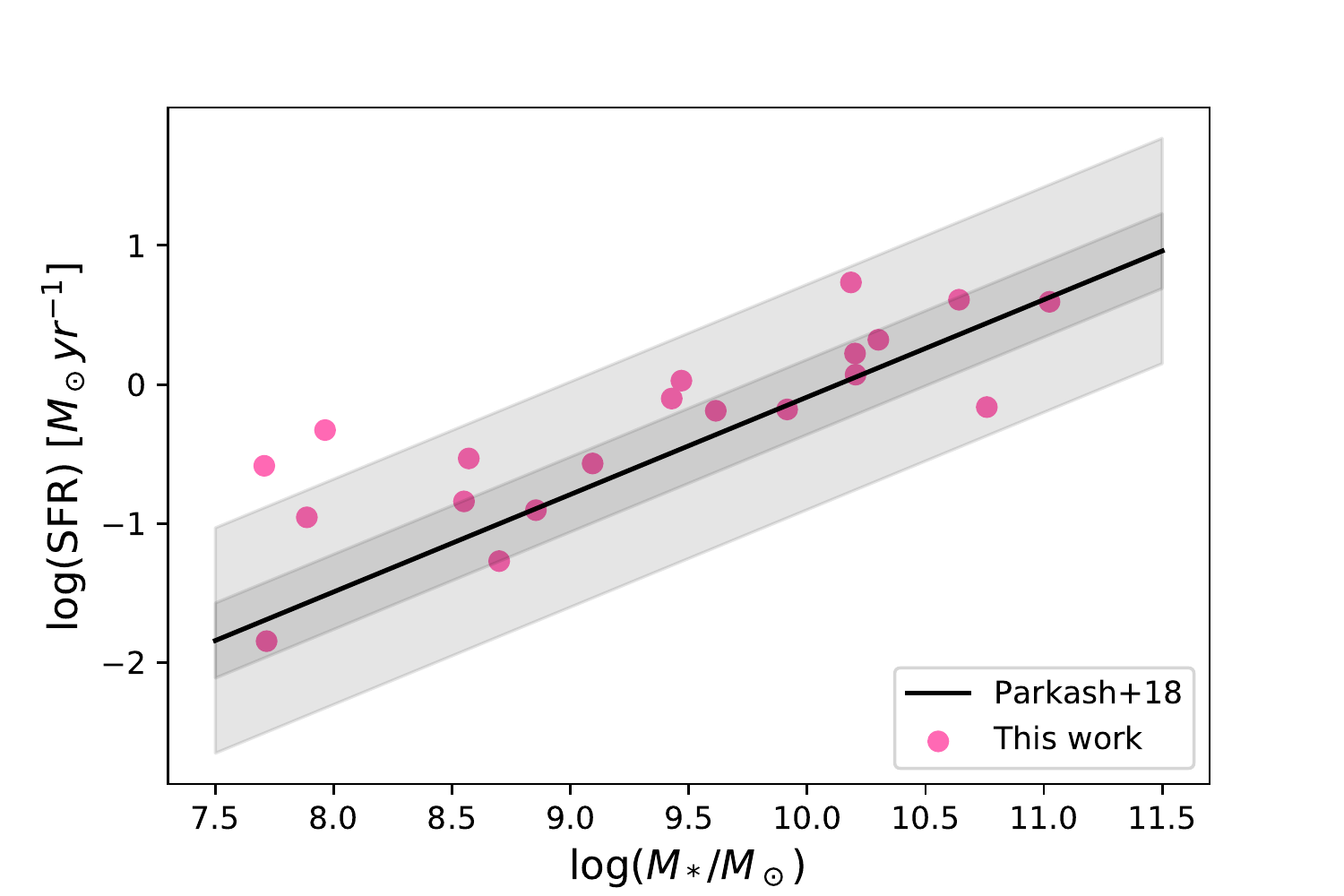}%
}
\caption{Left: The group members (green), compared to the $M_\mathrm{HI}-M_*$ relation from \citet{Maddox_2014} (black). Centre: The group members (red), compared to the $d-M_\mathrm{HI}$ relation (black) from \citet{Wang_2016}. Right: The group members (pink), compared to the main sequence of star-formation (black) from \citet{Parkash_2018}. The shaded region indicates the 1 and 3$\sigma$ scatter of the respective relations. Errors on all plots are relatively small compared to the size of the symbol.}
\label{fig:parkash}
\end{figure*}

\begin{figure*}
    \centering            
    \includegraphics[scale=0.62]{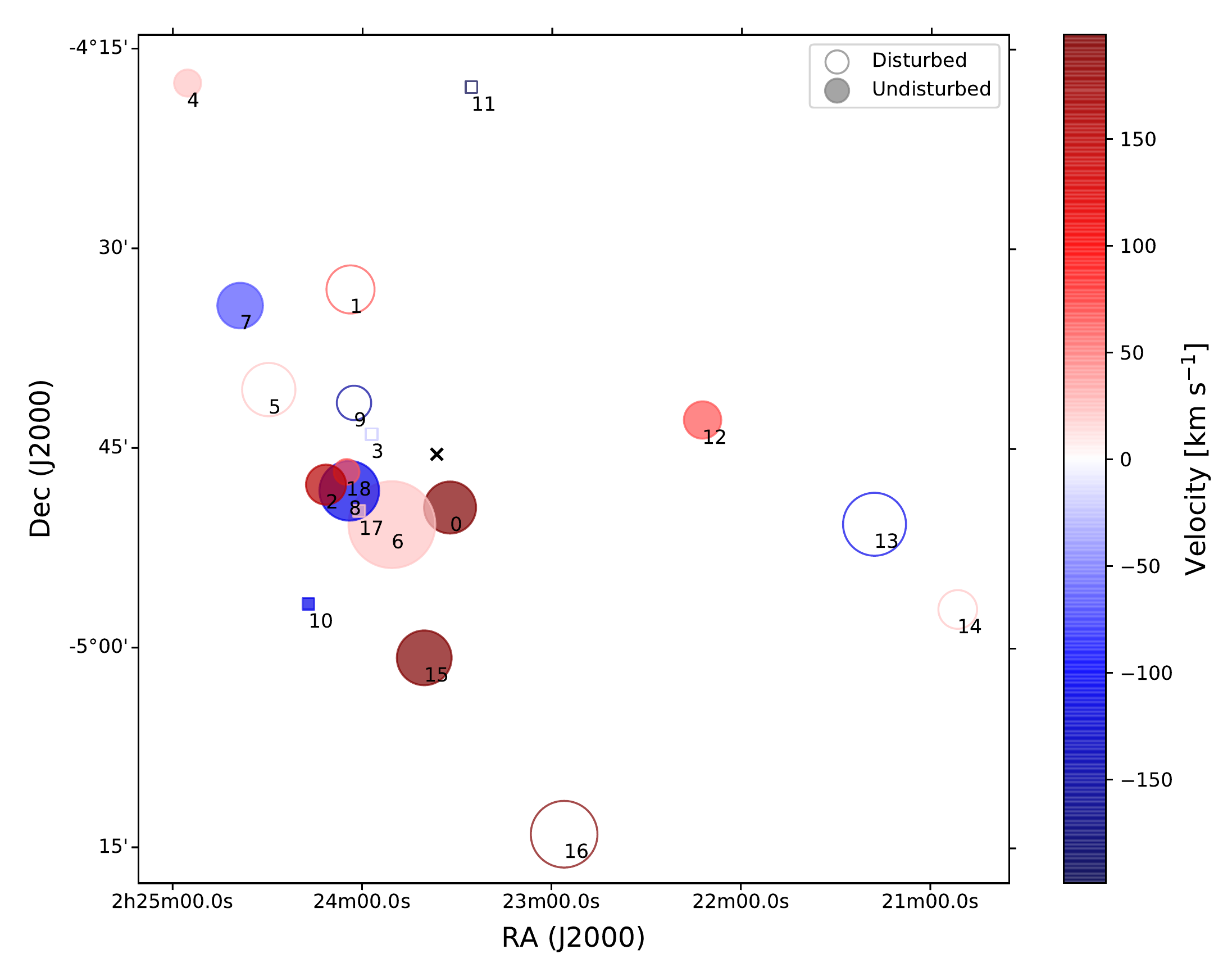}
    \caption{The position-velocity distribution of the 19 detected sources, classified by disturbances in their morphology. Sources without disturbed morphologies are shown as filled circles, and sources with disturbed morphologies are shown as unfilled circles. The size of the points represents an exponential scaling in $f^\mathrm{dyn}_\mathrm{HI}$ mass fraction, and sources without known dynamical masses are shown as squares. The colour of the points correspond to the relative systemic velocity of the individual galaxies ($\rm V_\mathrm{mean}=12,630\,$\kms). The black cross represents the unweighted average position of all 19 sources, which serves as an indicative group centre. }
    \label{fig:morphology}
\end{figure*}

We investigate the $M_\mathrm{HI}-M_*$ relation by comparing the galaxies in this group to the relation in \citet{Maddox_2014}, as shown in Fig.~\ref{fig:parkash}. All 19 sources fall within 3$\sigma$ of this relation, except for two low-\hi mass outliers. The outlier at $\log(M_*/\mathrm{M}_\odot)\sim 9.0$ corresponds to ID 18, which, as shown in Fig.~\ref{fig:3-mommap1}, has a low column density of $N_\mathrm{HI} \leq 2.9 \times 10^{20} \mathrm{cm}^{-2}$. There is likely fainter \hi emission below the sensitivity limit of this survey. Similarly, the outlier at $\log(M_*/\mathrm{M}_\odot)\sim 8.5$ corresponds to source ID 2.

We also compare our results to the diameter-\hi mass relation from \citet{Wang_2016}. This is shown in Fig.~\ref{fig:parkash}, where 15 of the 19 sources are plotted. The $d_\mathrm{HI}$ values here have been corrected for inclination, after which 4 sources had too low column densities to determine reliable measurements (Table~\ref{tab:results}). It should be noted that \citet{Wang_2016} assume $H_0 = 70$\kms Mpc$^{-1}$. This introduces a 6\% offset for diameter measurements, which is almost negligible for the range of values shown in Fig.~\ref{fig:parkash}. We choose to remain with the Planck 2018 cosmological parameters in order to keep consistency with the remainder of the MIGHTEE survey papers. All sources fall within the 3$\sigma$ confidence intervals of this relation, with the exception of source ID 5. This source has a long tidal tail. The disruption of the \hi has likely reduced the diameter of the disk.

To investigate whether the group galaxies are actively forming stars, we compare them to the star-formation main sequence (SFMS), which is the relation between stellar mass and SFR. Galaxies are considered to be quenched, or undergoing quenching if they reside below the relation, and are undergoing high star-formation if they reside above \citep{Kauffmann_2003}. In comparison to the SFMS from \citet{Parkash_2018}, we find that all \hi-detected galaxies, but two, lie within the $3\sigma$ scatter of the relation. This implies that they are actively undergoing star-formation. Two sources, ID's 12 and 16 are undergoing a high rate of star-formation. These sources have low masses in comparison to the rest of the sample, and both lie on the outskirts of the group.
%

\subsection{Morphological properties}
The sample includes some unusual and disturbed \hi morphologies. A special case is source ID 9, which shows an interacting pair of spiral galaxies, likely at the early stages of a potential merger. Comparable to the interacting galaxies in the M$\,81$ group \citep{Yun_1994}, in the optical, there are no tidal tails or other evidence that the galaxies are interacting, but there is a clear \hi bridge, indicating that they are indeed interacting. In the moment-1 map, there is a clear distinction between the two objects, a view that will be enhanced by the improved velocity resolution of future MIGHTEE observations. Interestingly, there is no obvious evidence of interaction in the integrated \hi spectrum of this source. This example highlights the importance of spatially-resolved \hi observations of environmental effects in groups and clusters. In particular, the spatial information provided by the MIGHTEE survey presents a view of a transformation process not evident in the integrated spectrum alone. 

Sources ID 1 and ID 5 also have somewhat distorted \hi distributions, suggestive of interactions with a putative intra-group medium. 
Source ID 5 is an edge-on spiral, with an off-centre \hi-disk. The disk extends further on the eastern side, with a large tidal tail on the northern end of the galaxy, in the opposite direction of the group's centre. The moment-0 map for source ID 1 shows a face-on disk and a clumpy, extended tail on the northern end. Sources 1D 14 and ID 16 also have disturbed \hi morphologies, but they lie towards the outskirts of the group. Both of them have an off-centre \hi disk. The disk asymmetry in these galaxies is perpendicular to the group centre. These sources are likely in the early stages of accreting into the galaxy group halo, resulting in tidal stripping from the host halo. 

A trend observed in most of the edge-on detected sources is an \hi deficiency in the centre of galaxies, as is frequently seen in spirals with large central bulges or high SFR \citep{Bigiel_2008}. This can be clearly seen in sources ID 3, ID 4, and ID 11 (Fig.~\ref{fig:3-mommap1}). There is a galaxy asymmetry or lopsidedness throughout the sample, particularly in the edge-on spiral galaxies. An example of this is the slightly disturbed morphology in source ID 11. On the southern side of the \hi disk, the \hi distribution appears to be flaring out towards the edges of the disk. Disturbed disks can also be seen in ID's 1, 3, 4, 7 and 9. Besides the bridge between the interacting pair, we detect no \hi residing outside of the detected galaxies, or other evidence of interaction between galaxies. 

Following \citet{Catinella_2013}, we expect groups with dynamical masses of $\log(M_\mathrm{dyn}/\mathrm{M}_\odot) < 13$ to have \hi-to-stellar mass fractions of $\log(f_\mathrm{HI}^\mathrm{*}) \sim -0.9$. We find a total $\log(f_\mathrm{HI}^\mathrm{*}) = -0.2$, which is 0.7 dex greater than expected. If no further stellar-mass dominant sources are present, then this group has an unusually large number of \hi-rich galaxies, which might suggest that it is in the early stages of assembly, despite being a relatively high-mass group. In addition to this, we do not see much evidence of galaxy-galaxy interaction, with the exception of ID 9, and the beginning of disturbed \hi disks, suggesting that there has not been much time for these interactions to take place. Furthermore, Fig.~\ref{fig:hist} shows no relation between stellar mass and relative velocity, implying that the most massive galaxies are not residing in the centre of the group, as expected for virialised groups. Finally, the velocity dispersion of the sources, shown in Fig.~\ref{fig:hist}, does not show the characteristic Gaussian distribution of a virialised group. This further suggests that the group is dynamically young.

To investigate the \hi morphologies in relation to their location in the group, we visually classified the galaxies into two broad classes based on the moment-0 maps, those with and without an obviously disturbed \hi morphology. This is admittedly a somewhat subjective classification scheme, particularly at this spatial and velocity resolution, and column density sensitivity. However, it provides an indicative guide in exploring how these objects may or may not be influenced by their environment, at least for the most obvious cases. This classification and the 3D positions of the sources in the group are plotted in Fig.~\ref{fig:morphology}. 
There also appears to be an asymmetry, showing some hints of more disturbed sources (ID's  1,  3,  5 and  9) just north of the apparent centre of the group. ID  3 has the lowest \hi-to-stellar mass ratio of the sample, implying that \hi has been removed from the system. This is consistent with \citet{Hess_2013}, who conclude that gas is preferentially removed from galaxies in the centre of groups. However, deeper observations and higher velocity resolution will be needed to make more definitive statements as the classification of several sources is ambiguous or marginal and potentially due to internal, secular processes. This ambiguity notwithstanding, the power of large samples of spatially resolved sources at a cosmologically significant distance is clearly evident in this serendipitously discovered \hi galaxy group. 

\section{Conclusion}\label{sec:conclusion}
We present an \hi-discovered galaxy group in the XMM-LSS field obtained from the MIGHTEE survey with the MeerKAT telescope. We detect 20 \hi galaxies (19 distinct \hi detections) down to a $3\sigma$ column density sensitivity of $N_\mathrm{HI} = 1.6\times10^{20}\,\mathrm{cm}^{-2}$, within an area of $2.3\deg^2$ and a recession velocity range of 12,679 -- 13,093 \kms ($z \sim 0.044$), and a spectral resolution of 46\kms. The 20 galaxies amount to a total group \hi mass of $\log(M_\mathrm{HI}/\mathrm{M}_\odot) = 11.20$. Using a simple model, we estimate a total group dynamical mass of $\log(M_\mathrm{dyn}/\mathrm{M}_\odot) = 12.32$.

The serendipitous discovery of this \hi group, in a well-studied field, highlights the importance of blind \hi surveys. \hi observations are a useful way to study and identify groups, and other loosely composed gravitationally bound systems at low redshift, that do not have detectable IGM haloes. The spatially resolved \hi emission has revealed a pair of interacting galaxies and numerous tidal tails that are not visible in optical images.

Overall, these results demonstrate the capability of MeerKAT with regard to \hi science, showing the advantages of its wide field-of-view coupled with high-fidelity, spatially-resolved imaging of \hi galaxies at cosmologically-significant distances. This galaxy group is an example of what is to come with the MIGHTEE survey and similar depth MeerKAT L-band observations: the discovery of several hundred \hi sources in a wide range of environments across the extragalactic legacy fields. The large cosmological volume probed by the MIGHTEE survey, and sensitivity of MeerKAT, will allow for the study of low-mass ($M_\mathrm{HI} \sim 10^7 \mathrm{M}_\odot$) galaxy populations in groups and other environments. This data will be significantly improved, with the currently ongoing finer velocity MIGHTEE observations ($\sim5$\kms), allowing for more accurate kinematic modelling of the resolved galaxies.

\section*{Acknowledgements}

We thank the anonymous referee for their quick and helpful comments. We thank Thijs van der Hulst for helpful comments which improved this paper. The research of SR and RPD is supported by the South African Research Chairs Initiative (grant ID 77948) of the Department of Science and Innovation and National Research Foundation. SR, RPD and MJJ acknowledge the financial assistance of the South African Radio Astronomy Observatory (SARAO) towards this research (www.ska.ac.za). The MeerKAT telescope is operated by the South African Radio Astronomy Observatory, which is a facility of the National Research Foundation, an agency of the Department of Science and Innovation. We acknowledge use of the Inter-University Institute for Data Intensive Astronomy (IDIA) data intensive research cloud for data processing. IDIA is a South African university partnership involving the University of Cape Town, the University of Pretoria and the University of the Western Cape. The authors acknowledge the Centre for High Performance Computing (CHPC), South Africa, for providing computational resources to this research project. This work has made use of the Cube Analysis and Rendering Tool for Astronomy \citep[CARTA][]{Comrie_2021}. This work is based on data products from observations made with ESO Telescopes at the La Silla Paranal Observatory under ESO programme ID 179.A-2005 (Ultra-VISTA) and ID 179.A- 2006(VIDEO) and on data products produced by CALET and the Cambridge Astronomy Survey Unit on behalf of the Ultra-VISTA and VIDEO consortia. Based on observations obtained with MegaPrime/MegaCam, a joint project of CFHT and CEA/IRFU, at the Canada-France-Hawaii Telescope (CFHT) which is operated by the National Research Council (NRC) of Canada, the Institut National des Science de l’Univers of the Centre National de la Recherche Scientifique (CNRS) of France, and the University of Hawaii. This work is based in part on data products produced at Terapix available at the Canadian Astronomy Data Centre as part of the Canada-France-Hawaii Telescope Legacy Survey, a collaborative project of NRC and CNRS.The Hyper Suprime-Cam (HSC) collaboration includes the astronomical communities of Japan and Taiwan, and Princeton University. The HSC instrumentation and software were developed by the National Astronomical Observatory of Japan (NAOJ), the Kavli Institute for the Physics and Mathematics of the Universe (Kavli IPMU), the University of Tokyo, the High Energy Accelerator Research Organization (KEK), the Academia Sinica Institute for Astronomy and Astrophysics in Taiwan (ASIAA), and Princeton University. Funding was contributed by the FIRST program from Japanese Cabinet Office, the Ministry of Education, Culture, Sports, Science and Technology (MEXT), the Japan Society for the Promotion of Science (JSPS), Japan Science and Technology Agency (JST), the Toray Science Foundation, NAOJ, Kavli IPMU, KEK, ASIAA, and Princeton University. AAP acknowledges the support of the STFC consolidated grant ST/S000488/1. WM is supported by the South African Research Chairs Initiative of the Department of Science and Technology and National Research Foundation. RB acknowledges support from the Glasstone Foundation and from an STFC Ernest Rutherford Fellowship [grant number ST/T003596/1]. NM acknowledges support from the Bundesministerium f{\"u}r Bildung und Forschung (BMBF) award 05A20WM4. MJJ acknowledges  support from the UK Science and Technology Facilities Council [ST/S000488/1 and ST/N000919/1]. MT and MJJ acknowledge support from the Oxford Hintze Centre for Astrophysical Surveys which is funded through generous support from the Hintze Family Charitable Foundation. LVM acknowledges financial support from the grants AYA2015-65973-C3-1-R and RTI2018-096228-B-C31 (MINECO/FEDER, UE), as well as from the State Agency for Research of the Spanish MCIU through the “Center of Excellence Severo Ochoa” award to the Instituto de Astrof{\'i}sica de Andaluc{\'i}a (SEV-2017-0709). This research made use of Astropy,\footnote{http://www.astropy.org} a community-developed core Python package for Astronomy. MG and JC acknowledge financial support from the Inter-University Institute for Data Intensive Astronomy (IDIA).

\section*{Data Availability}

The data underlying this article are subject to the proprietary period and data distribution policies of the MIGHTEE Large Survey Project. 
{ The MIGHTEE-\hi spectral cubes will be released as part of the first data release of the MIGHTEE survey, which will include cubelets of the sources discussed in this paper. The derived quantities from the multi-wavelength ancillary data will be released with the final data release of the VIDEO survey in mid-2021. Alternative products are already available from the Herschel Extragalactic Legacy Project \citep[HELP;][]{Shirley_2021} and also soon from the Deep Extragalactic VIsible Legacy Survey \citep[DEVILS;][]{davies2021}.}



\bibliographystyle{mnras}
\bibliography{example} 





\bsp	
\label{lastpage}
\end{document}